\documentclass[letterpaper,twocolumn,prx,aps,superscriptaddress,floatfix,longbibliography]{revtex4-2}
\usepackage[latin1]{inputenc}
\usepackage[T1]{fontenc}
\usepackage{bm}
\usepackage{amssymb,amsbsy,amsmath}
\usepackage{graphicx}
\usepackage{verbatim}
\usepackage{color}
\usepackage{pifont}
\usepackage{hyperref}
\usepackage[dvipsnames]{xcolor}
\usepackage{verbatim}
\usepackage{esvect}
\usepackage{float}
\usepackage{algorithm,algpseudocode,float}
\usepackage{lipsum}

\makeatletter

\makeatletter

\renewenvironment{widetext@grid}{%
  \par\ignorespaces
  \setbox\widetext@top\vbox{%
   \vskip15\p@
   \hb@xt@\hsize{%
    \leaders\hrule\hfil
    \vrule\@height6\p@
   }%
   \vskip6\p@
  }%
  \setbox\widetext@bot\hb@xt@\hsize{%
    \vrule\@depth6\p@
    \leaders\hrule\hfil
  }%
  \onecolumngrid
  \let\set@footnotewidth\set@footnotewidth@ii
}{%
  \par
  \twocolumngrid\global\@ignoretrue
  \@endpetrue
}%

\makeatother

\hypersetup{
	colorlinks=true,       		
	linkcolor=orange,          	
	citecolor=blue,             
	urlcolor=blue,          
}
\usepackage{braket}

\begin{document}

	\title{Versatile photonic frequency synthetic dimensions using a single programmable on-chip device}

    \author{Zhao-An~Wang}
    \thanks{These authors contributed equally to this work.}
    \author{Xiao-Dong~Zeng}
    \thanks{These authors contributed equally to this work.}
    \author{Yi-Tao~Wang}
    \thanks{These authors contributed equally to this work.}
    \author{Jia-Ming~Ren}
    \author{Chun~Ao}
    \author{Zhi-Peng Li}
    \author{Wei~Liu}
    \affiliation{CAS Key Laboratory of Quantum Information, University of Science and Technology of China, Hefei, 230026, China}
    \affiliation{Anhui Province Key Laboratory of Quantum Network, University of Science and Technology of China, Hefei, Anhui 230026, China}
    \affiliation{CAS Center For Excellence in Quantum Information and Quantum Physics, University of Science and Technology of China, Hefei, 230026, China}
    \author{Nai-Jie~Guo}
    \affiliation{CAS Key Laboratory of Quantum Information, University of Science and Technology of China, Hefei, 230026, China}
    \affiliation{Anhui Province Key Laboratory of Quantum Network, University of Science and Technology of China, Hefei, Anhui 230026, China}
    \affiliation{CAS Center For Excellence in Quantum Information and Quantum Physics, University of Science and Technology of China, Hefei, 230026, China}
    \affiliation{Hefei National Laboratory, University of Science and Technology of China, Hefei 230088, China}
    \author{Lin-Ke~Xie}
    \affiliation{CAS Key Laboratory of Quantum Information, University of Science and Technology of China, Hefei, 230026, China}
    \affiliation{Anhui Province Key Laboratory of Quantum Network, University of Science and Technology of China, Hefei, Anhui 230026, China}
    \affiliation{CAS Center For Excellence in Quantum Information and Quantum Physics, University of Science and Technology of China, Hefei, 230026, China}
    \author{Jun-You~Liu}
    \affiliation{CAS Key Laboratory of Quantum Information, University of Science and Technology of China, Hefei, 230026, China}
    \affiliation{Anhui Province Key Laboratory of Quantum Network, University of Science and Technology of China, Hefei, Anhui 230026, China}
    \affiliation{CAS Center For Excellence in Quantum Information and Quantum Physics, University of Science and Technology of China, Hefei, 230026, China}
    \affiliation{Hefei National Laboratory, University of Science and Technology of China, Hefei 230088, China}
    \author{Yu-Hang~Ma}
    \author{Ya-Qi~Wu}
    \affiliation{CAS Key Laboratory of Quantum Information, University of Science and Technology of China, Hefei, 230026, China}
    \affiliation{Anhui Province Key Laboratory of Quantum Network, University of Science and Technology of China, Hefei, Anhui 230026, China}
    \affiliation{CAS Center For Excellence in Quantum Information and Quantum Physics, University of Science and Technology of China, Hefei, 230026, China}
    \author{Xi-Wang~Luo}
    \author{Shuang~Wang}
    \author{Jian-Shun~Tang}
    \email{tjs@ustc.edu.cn}
    \author{Chuan-Feng~Li}
    \email{cfli@ustc.edu.cn}
    \author{Guang-Can~Guo}
    \affiliation{CAS Key Laboratory of Quantum Information, University of Science and Technology of China, Hefei, 230026, China}
    \affiliation{Anhui Province Key Laboratory of Quantum Network, University of Science and Technology of China, Hefei, Anhui 230026, China}
    \affiliation{CAS Center For Excellence in Quantum Information and Quantum Physics, University of Science and Technology of China, Hefei, 230026, China}
    \affiliation{Hefei National Laboratory, University of Science and Technology of China, Hefei 230088, China}

	
	\renewcommand{\figurename}{Fig.}
	
	\newcommand{\Todos}[1]{\textcolor{red}{#1}}
			
\begin{abstract}

Investigating physical models with photonic synthetic dimensions has been generating great interest in vast fields of science. The rapidly developing thin-film lithium niobate (TFLN) platform, for its numerous advantages including high electro-optic coefficient and scalability, is well compatible with the realization of synthetic dimensions in the frequency together with spatial domain. While coupling resonators with fixed beam splitters is a common experimental approach, it often lacks tunability and limits coupling between adjacent lattices to sites occupying the same frequency domain positions. Here, on the contrary, we conceive the resonator arrays connected by electro-optic tunable Mach-Zehnder interferometers in our configuration instead of fixed beam splitters. By applying bias voltage and RF modulation on the interferometers, our design extends such coupling to long-range scenario and allows for continuous tuning on each coupling strength and synthetic effective magnetic flux. Therefore, our design enriches controllable coupling types that are essential for building programmable lattice networks and significantly increases versatility. As the example, we experimentally fabricate a two-resonator prototype on the TFLN platform, and on this single chip we realize well-known models including tight-binding lattices, the Hall ladder and Creutz ladder. We directly observe the band structures in the quasi-momentum space and important phenomena such as spin-momentum locking, flat band and the Aharonov-Bohm cage effect. These results demonstrate the potential for convenient simulations of more complex models in our configuration.

\end{abstract}

\maketitle

\noindent\large\textbf{Introduction}

\begin{figure*}[t]
    \centering
    \includegraphics[width=0.9\textwidth]{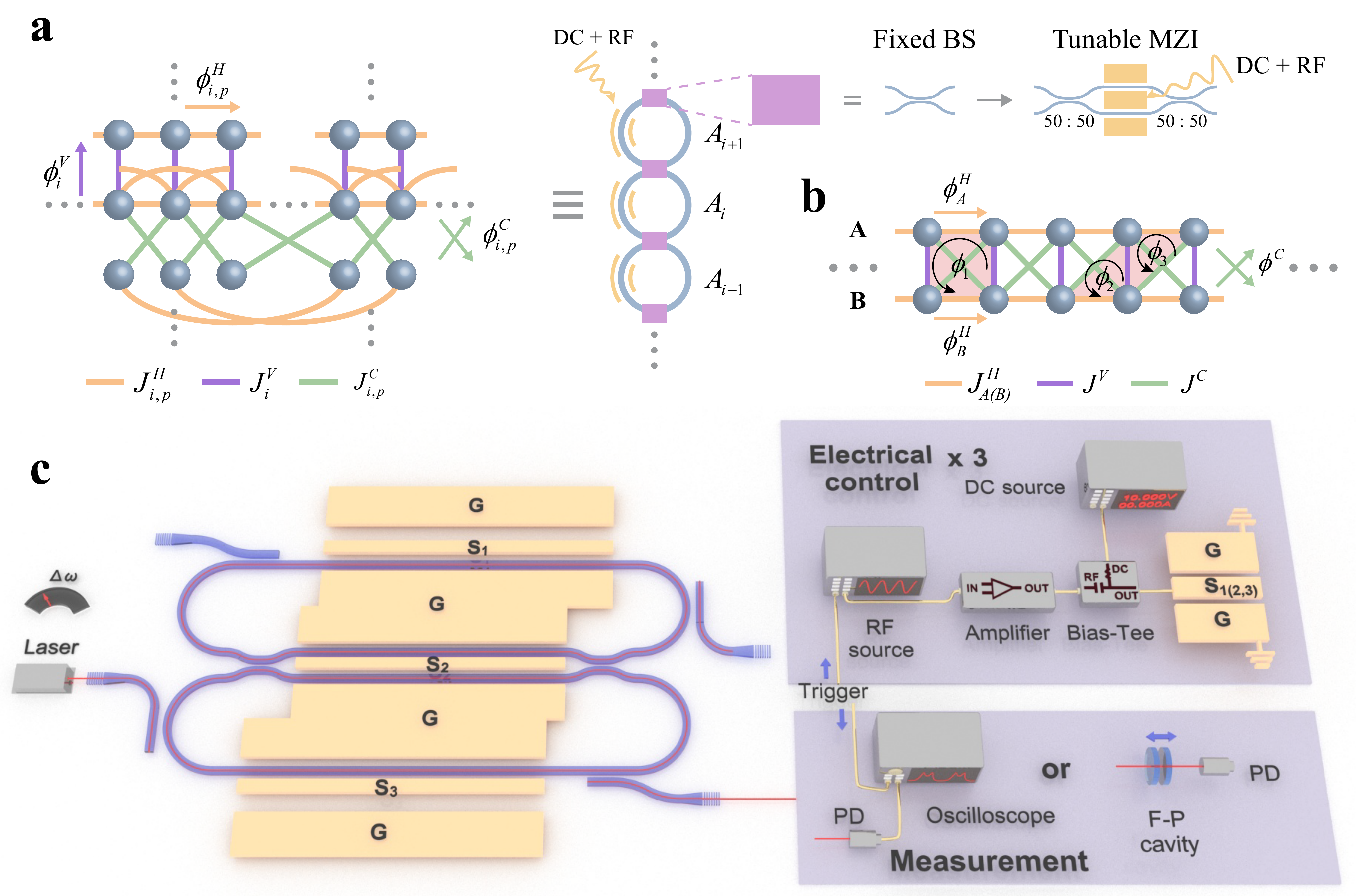}
    \caption{\textbf{Schematic and experimental setup.} (\textbf{a}) Configuration of a lattice network in frequency synthetic dimensions. The lattice network is simulated by a train of modulated resonators. Instead of fixed beam splitters (BS), MZIs tunable by DC and RF signals are used to couple the adjacent resonators (purple boxes). The lattice network consequently contains three types of coupling $J^H_{i,p}$ (orange), $J^V_i$ (violet), $J^C_{i,p}$ (green), where $i$ is the index of resonators ($A_i$) along the spatial direction, and $p$ is the coupling range along the frequency direction. These couplings stem from RF signals on the resonators, DC and RF signals applied on the MZIs, respectively. $\phi^H_{i,p}$, $\phi^V_{i}$, $\phi^C_{i,p}$ are the corresponding phases accumulated while coupling, which stem from the initial phases of the modulation. (\textbf{b}) Schematic of the Creutz ladder consisting of two lattices A and B which serve as two pseudospins. The coupling types are of same definition in \textbf{a} and are abbreviated as $J^H_{A(B)}$, $J^V$ and $J^C$. The phases accumulated around the rectangular or triangular plaquettes form gauge potentials such as $\phi_1$, $\phi_2$ and $\phi_3$. (\textbf{c}) Experimental setup for realizing \textbf{b} with two MZI assisted race-track resonators on TFLN. Three sets of electrical signals consisting of DC and RF parts are applied on three sets of the ground-signal-ground (GSG) arranged electrodes through Bias-Tees to control the two resonators ($S_1,S_3$) and the MZI ($S_2$). For detecting band structures, a probe laser with detuning $\Delta\omega$ is injected to excite the bands. The time- (quasi-momentum-) varying transmittance signal, which is a slice of the band structure, is detected by a photodetector (PD) followed by an oscilloscope triggered by the RF source. For obtaining mode distribution, the measurement part is replaced by scanning a Fabry-Perot (F-P) cavity together with a photodetector.}
    \label{fig:scheme}
\end{figure*}

\normalsize
Creating a controllable artificial system, of which the dynamics highly resembles the one in real space, is a commonly used method to simulate inaccessible, complex and high-dimensional systems. The artificial systems also called synthetic dimensions, are therefore capable to investigate fascinating novel physics in the field of such as solid-state physics \cite{lustig2019,ozawa2019,TopoRev,sridhar2024}. In terms of realization, photonic systems hold advantages since it has multiple manipulable degrees of freedom including polarization, space, frequency and orbital angular momentum \cite{regensburger2012,Luo2015,luo2017,Yuan18,Yuan2019,Yuan2021,YuanPRB,Fang2012,lumer2019,wang2020,chen2021,li2023}. Configuring synthetic dimensions in the form of optical frequency lattices is one of the popular approaches in pioneering works, where optical-fiber-based resonators are mainly exploited \cite{Dutt2019,Dutt2020,Dutt2022,Li2021,WangK2021,li2023,wang2021}. 

Apart from optical fibers, integrated optics holds promise for generating even more complex and programmable lattice networks for its good scalability, stability and reconfigurablity. Especially, with broad transparent spectrum, high non-linear and electro-optic coefficients, thin-film lithium niobate (TFLN) provides an ideal platform. The rapid development of high performance devices on such platform including modulated resonators have paved the way for frequency synthetic dimensions \cite{LNRev1,LNRev2,zhang2019,Zhang2017,feng2024,yu2022,xu2020,Ozawa2016,Lieb}. As a result, constructing frequency lattices on TFLN has garnered increasing interest recently and several experimental works, such as Refs. \cite{Dinh2024,ye2024,Javid2023,Hu2020,hu2022} and our previous work Ref.\,\cite{wza2024}, have been reported.

To advance synthetic dimensions in lattices and interactions, realizing coherent coupling between resonators is crucial. When two resonators are coupled via a passive coupler, adjacent coupling arises between lattice points (resonators) at identical frequencies. This approach facilitates the combination of frequency and spatial dimensions, enabling the simulation of diverse models and intriguing phenomena such as the Hall ladder, the Lieb lattice, band structures with effective magnetic fields, and spin-momentum locking, among others \cite{Dutt2020,Lieb,hugel2014}. Furthermore, controllable long-range coupling, particularly cross coupling, is essential for realizing broader classes of models, for example, a historically significant example is the Creutz ladder, one of the earliest models of chiral lattice fermions. This model features edge states protected by spatial inversion symmetry and is characterized by a $\mathbb{Z}$-type topological invariant, even in the absence of chiral symmetry for most cases \cite{creutz1999,zurita2020,Hung2021,He2021}. Achieving cross coupling requires linking lattice points with distinct frequencies and spatial positions. However, experimental demonstration of such coupling within a frequency-spatial hybrid configuration, employing sufficiently tunable methods, remains an open challenge. Therefore, enhancing the tunability of coupling modes between resonators is necessary to engineer lattice couplings and realize a wider range of physical models.

In this work, we design and fabricate a Mach-Zehnder-interferometer (MZI)-assisted double-resonator device on TFLN, and then realize the sufficient regulation of coupling strength and range in synthetic dimensions. By tuning the bias voltage applied on the MZI \cite{dai2024}, we are able to arbitrarily tune the same-frequency coupling strength between two resonators. More importantly, by modulating the MZI rather than resonators with RF signals, we further extend the adjacent coupling to cross coupling and long-range coupling, without introducing auxiliary resonators \cite{YuanPRB,lin2018}. This enable us to simulate more complex models like the Creutz ladder. We show the experimental observation of band structures of tight-binding lattices, the Hall ladder and the Creutz ladder, equipped with gauge potentials. Moreover, we also directly observe topological features including the Aharonov-Bohm cage effect which was first carried out by Vidal \textit{et al} \cite{Vidal1998,Vidal2000,Vidal2001,Benoit2002}. While maintaining a similar scale and fabrication difficulty to using fixed beam splitters as couplers, our configuration significantly increases the versatility of devices for realizing frequency synthetic dimensions and broadens the scope of models that can be simulated.

\noindent\large\textbf{Results}


\begin{figure*}
    \centering
    \includegraphics[width=0.95
\textwidth]{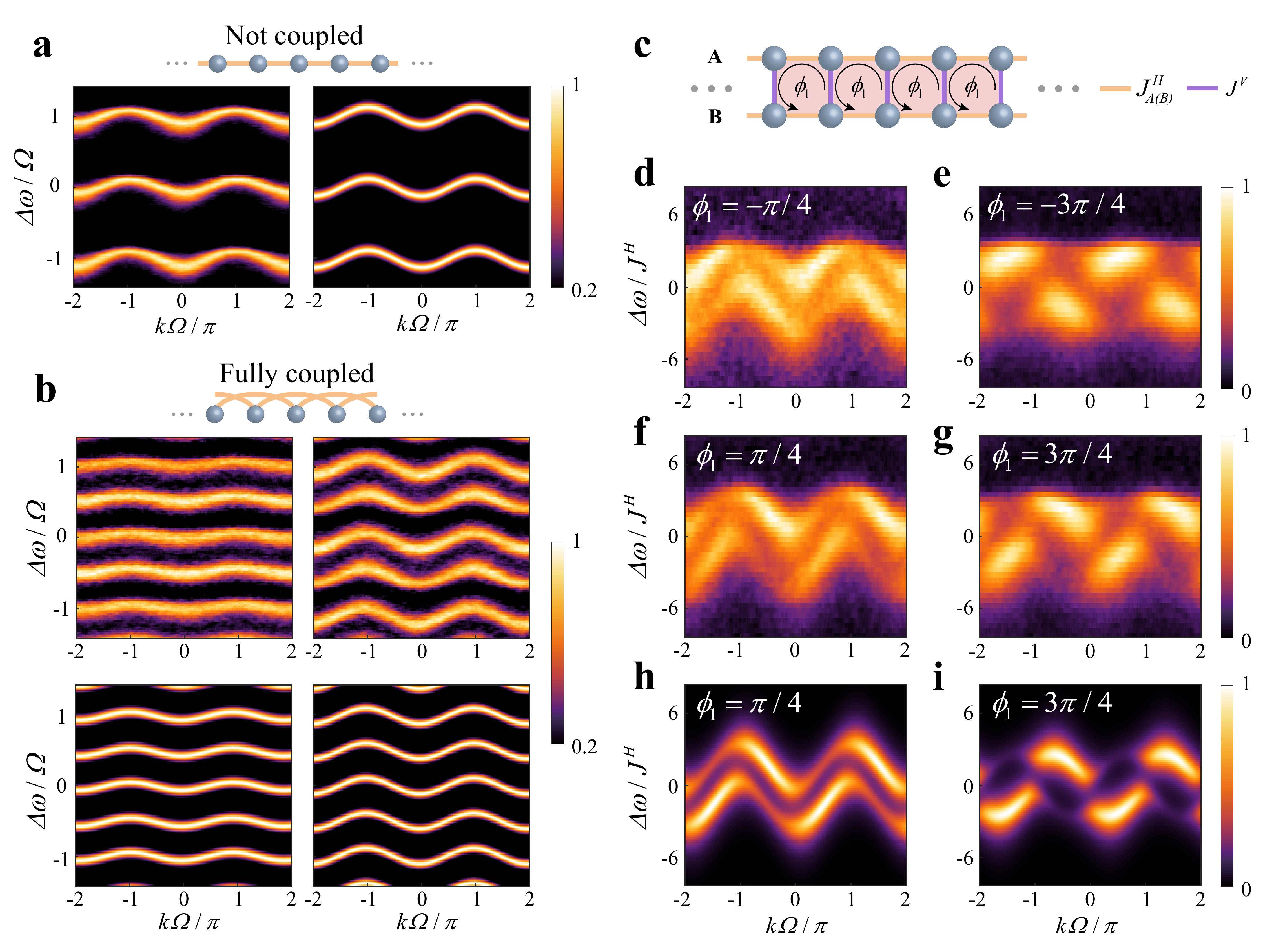}
    \caption{\textbf{Experimental obtained band structures of tight-binding lattices and the Hall ladder.} The band structures exhibit tight-binding lattice characters when two resonators are either not coupled ($J^V=0$, \textbf{a}) or fully coupled into a large resonator whose length is doubled and FSR is halved (\textbf{b}). $J_B^H$ is set to zero in the left panels of (\textbf{b}), while set to be equal to $J_A^H$ in the right panels. The right panel in (\textbf{a}) and the lower panels in (\textbf{b}) are the corresponding numerical calculations. (\textbf{c}) Illustration of the Hall ladder where the crossing coupling is absent. (\textbf{d} to \textbf{g}) The heat maps are the experimentally obtained band structures of the Hall ladder with coupling strength ratio $r_{VH}^{}=J_V/2J_H=0.75$, $J^H=0.06\Omega$ and an effective magnetic flux $\phi_1$ being $\pm \pi/4$ or $\pm 3\pi/4$. (\textbf{h} and \textbf{i}) are the numerical calculations of (\textbf{f} and \textbf{g}). The population information on the two spins (resonators) are encoded in the normalized intensities. Choosing the opposite flux ($\phi_1$) is equivalent to selectively exciting the opposite spin. For one spin (\textbf{d} and \textbf{e}), the negative (positive) $k$ states in the first Brillouin zone predominate the upper (lower) bands. Meanwhile, complementary patterns appear in the other spin (\textbf{f} and \textbf{g}), signifying the spin-momentum locking.}
    \label{fig:Hall}
\end{figure*}

\noindent\normalsize\textbf{Theoretical framework} 

\normalsize Design of the couplings among the lattice network points is one of the core ingredients of the configuration of physical models, specifically, both the coupled modes and coupling range are required to be tuned appropriately. Towards an ideal versatile and programmable network, a device being capable to cover as many coupling types as possible is on demand. Now we consider a 2D network protocol constituted by a series of identical adjacently-coupled 1D lattices, which can be simulated by a train of resonators (Fig.\,\ref{fig:scheme}a), by integrating the frequency and the spatial domain. 
We can depict the general model with the horizontal, vertical, crossing coupling strengths ($J^H_{i,p}$, $J^V_{i}$, $J^{C}_{i,p}$), as well as the corresponding coupling phases ($\phi_{i,p}^H$, $\phi_{p}^V$, $\phi_{i,p}^{C}$), where $i$ denotes the $i$-th lattice. These phases form effective magnetic fluxes threading the plaquettes of the models, introducing gauge potentials. In the frequency domain, the synthetic lattice points are formed by the resonant modes of each resonator with identical spacing of the free spectral range (FSR) $\Omega$. The phase modulation on the resonators with frequency equal to $p\Omega$ couples the $n$-th and ($n+p$)-th lattice points of each lattice (resonator), signifying the horizontal coupling $J^H_{i,p}$. The vertical coupling $J^V_i$ indicates the interaction purely in the spatial domain, i.e., the coupling between the nearest inter-lattice points of the adjacent lattices, which was commonly achieved by static coupling of two resonators using a beam splitter with a fixed splitting ratio. 

To promote convenience and versatility, we replace the splitter with a electro-opitc tunable $2\times2$ MZI (Fig.\,\ref{fig:scheme}a) which has two fold of functions. The static coupling is responsible for $J^V$. As the splitting ratio of the MZI, standing for static coupling, can be continuously tuned from zero to unity with a direct-current (DC) voltage, the lattices can be arbitrarily adjusted from being statically completely uncoupled ($J^V=0$) to being fully coupled into a lattice with spacing $\Omega/2$. This advantage can particularly benefit the applications where fine coupling-strength tuning are required. 

\begin{figure*}[t]
    \centering
    \includegraphics[width=\textwidth]{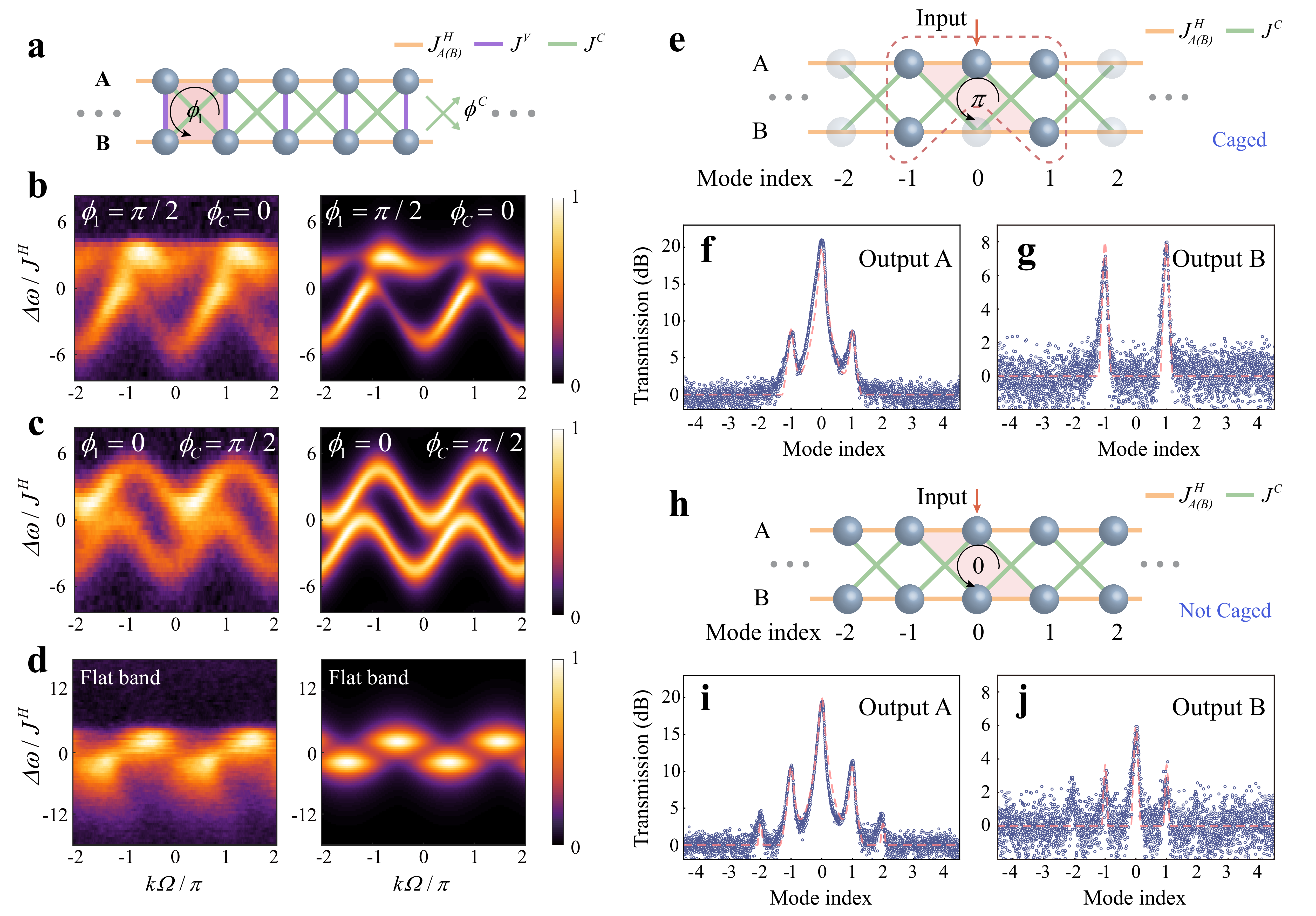}
    \caption{\textbf{Experimental obtained band structures of the Creutz ladder and direct observation of the Aharonov-Bohm cage effect.} (a) Illustration of the Creutz ladder. (\textbf{b} and \textbf{c}) The heat maps display two general band structures of the Creutz ladder given different $\phi_1$ and $\phi^C$, where $J^V/2J^H=1.1$, $J^C/J^H=0.52$ and $J^H=0.06\Omega$. (\textbf{d}) The flat band structure measured at $J^V=0$, $J^H=J^C=0.028\Omega$ and $-\phi_A^{H}=\phi_B^H=\pi/2$. The absence of $J^V$, equal $J^H$ and $J^C$, together with the $\pi$ flux also satisfy the Aharonov-Bohm cage effect condition. The left panels in (\textbf{b} to \textbf{d}) are the experimental results and the right panels are the numerical calculations. The population information on the two spins (resonators) are encoded in the normalized intensities. (\textbf{e}) Illustration of the Aharonov-Bohm cage effect where the phase collected in a round trip in the pink area is $\pi$. When a probe light is input from resonator A with frequency near the zero index mode, the distribution is caged within the dashed box ($0, \pm 1$ in lattice A and $\pm 1$ in lattice B). (\textbf{f} and \textbf{g}) The blue points are the experimental readout of the mode distribution from the drop ports of resonator A and B measured by a Fabry-Perot (F-P) cavity. The pink dashed lines represent the normalized numerical calculations, where $\Delta\omega$ is fitted to be approximate $4J^H$. The distribution other than those inside the dashed box is clearly suppressed. (\textbf{h}) Illustration of the not caged situation where the phase collected in the pink area is zero. The light normally spreads out along the frequency direction. (\textbf{i} and \textbf{j}) The blue points are experimental results of the not caged situation, where more modes survives compared to the caged ones. The pink circles and dashed lines represent the normalized numerical calculations, where $\Delta\omega$ is fitted to be approximate $2J^H$. The distribution results are plotted in the logarithmic coordinates and the linewidth of the F-P cavity is taken into account.}
    \label{fig:Creutz}
\end{figure*}
Moreover, we highlight that this design makes it possible to deflect the interaction along the spatial direction towards the frequency direction---the modulation of the coupling of two resonators is responsible for $J^C_{i,p}$. To be concrete, by adding modulation of frequency $p\Omega$ on the MZI other than the resonators, the mode hops $p$ sites along the frequency direction while crossing to the adjacent resonators, which manifests the cross coupling is satisfied. The introducing of long-range interaction between lattices remarkably raise the complexity of the model and provide opportunity to directly modify the off-diagonal elements of the Hamiltonian matrix in the quasimomentum space.

Without losing generality, we show experimental demonstration of the fundamental ingredient of the network --- the two-resonator protocol in the following. We denote the resonators by A and B, respectively. The protocol is capable of simulating 1D tight-binding lattice, as well as quasi-1D two-legged ladder models where the legs (resonators) represent two pseudospins ($s_+$, $s_-$), including the Hall ladder and the Creutz ladder. Varying fluxes in ladder systems alters the Hamiltonians, leading to consequences such as time-reversal symmetry breaking, changes of the population information in band structures, interference in hopping processes, and possible potential topological phase transitions. These ladders are one of the simplest models while retaining important properties such as spin-orbit coupling and band flatness. Retaining the nearest neighbouring coupling in the horizontal and crossing coupling ($p=1$) and setting $\phi^V$ to zero, the model turns into the Creutz ladder. In the standard Creutz ladder, the parameters are evenly set as $J_A^H=J_B^H=J^C$, $\phi_A^H=-\phi_B^H$ and $\phi^C=0$; and also in the standard Hall ladder, $J_A^H=J_B^H$ and $\phi_A^H=-\phi_B^H$. Here, we still call our models the Creutz ladder or the Hall ladder according to their shapes for simplicity, although our models are more general which can as well include the uneven cases. We show its schematic in Fig.\,\ref{fig:scheme}b, of which the Hamiltonian in the real space and quasi-momentum ($k$) space reads

\begin{equation}
\begin{aligned}
    &H = -\sum_{n} (J^H_{A} e^{-i \phi_{A}^H} a_{n+1}^\dagger a_{n}^{} + J^H_{B} e^{-i \phi_{B}^H} b_{n+1}^\dagger b_{n}^{}) \\
    &- \sum_{n} J^V b_{n}^\dagger a_{n}^{} - \sum_{n} J^{C} e^{-i \phi^C} (a_{n+1}^\dagger b_{n}^{} + b_{n+1}^\dagger a_{n}^{})+ h.c., \\
        \\
    &H_k=- 2 J^H_{A} \cos(k\Omega+\phi_A^H) \sigma_+ - 2 J^H_{B} \cos(k\Omega+\phi_B^H) \sigma_- \\
    &- [J^V+2 J^C \cos{(k\Omega+\phi_C)}]\sigma_x,
    \label{eq.exp}
\end{aligned} 
\end{equation}
where \textit{h.c.} representing the Hermitian conjugate, $\sigma_x$, $\sigma_y$, $\sigma_z$ and $\sigma_0$ are Pauli operators on pseudospins, and $\sigma_\pm=(\sigma_0\pm\sigma_z)/2$.

\noindent\normalsize\textbf{Experiment}

\normalsize The experimental setup is illustrated in Fig.\,\ref{fig:scheme}c. On a TFLN platform, we design and fabricate the double resonators coupled by a MZI with three sets of electrodes. The DC signal applied on the resonators through the Bias-Tees is to align their resonant frequencies. Given the reciprocal relationship between the time domain and the frequency domain, time can be considered as the quasi-momentum ($k$) associated with the frequency lattice. We obtain the $k$-space band structures with the time-resolved spectroscopy approach that is proposed in Ref.\,\cite{Dutt2019}. This can be analyzed utilizing Floquet theory that may be extended beyond the rotating-wave approximation (RWA, but our experiment is typically within the RWA) \cite{Dutt2019,Dutt2020}, or scattering matrix approach\cite{cheng2025}. The derivation can be found in Supplementary Information. Adjusting the detuning $\Delta\omega$ of the probe laser injected into resonator A, we can selectively excite different slices of the band structures. The data is collected from the drop port of the same resonator, which leads to the projection of the band structures on the lattice A. 
The time-resolved data for each excited slice enables the reconstruction of the complete projected band structure via data stacking. The pattern intensities are proportional to the square of the population on lattice A. Additionally, a piezoelectrically-driven Fabry-Perot cavity serves as an optical spectrum analyzer to filter specific frequency components from the time-resolved data, enabling detailed analysis of the mode distribution.

\textbf{Tight-binding single lattice and Hall ladder.} We start from applying local modulation on the resonators and only DC signals on the MZI ($J^C=0$), reducing the Creutz ladder to the Hall ladder. Specially, when the MZI is tuned to disconnect two resonators or to fully couple them, the model further reduces to a tight-binding single lattice. 

In Fig.\,\ref{fig:Hall}, a and b, we present the experimentally observed sinusoidal band structures of the tight-binding single lattices. On one hand, when the resonators are not coupled, the lattice constant is $\Omega$ and the nearest-neighbouring lattice points are coupled. On the other hand, when the resonators are fully coupled, the effective length of the ring is doubled, halving the FSR (lattice constant). The modulation is kept equal to the original FSR and the next-near-neighbouring lattice points are coupled.

Fig.\,\ref{fig:Hall}, d-i, illustrates the projected band structures of the Hall ladder with various synthetic magnetic flux $\phi_1=-2\phi_A^{H}=2\phi_B^{H}$ that indicates gauge potentials. As the two resonators can be depicted as pseudospins ($s_+$, $s_-$), the Hall ladder can signify spin-orbit coupling controlled by the magnetic field. 
Let $J^{H}=J^{H}_{A}=J^{H}_{B}$ and with the proper choice of $\phi_1$, we observe a pronounced dependence of the pseudospin character on $k$. The probe laser selectively excites the band associated with pseudospin $s_+$, represented by the incident resonator A (Fig.\,\ref{fig:Hall}, d and e). In the first Brillouin zone, for $k>0$, the excitation occurs in the upper branch,  whereas for $k<0$, it occurs in the lower branch.  Upon inverting the magnetic flux to  $-\phi_1$, resonator A turns into pseudospin $n_-$ due to symmetry. Consequently, the band structures exhibit mirrored patterns, indicative of chiral spin-momentum locking.

\textbf{Creutz ladder.} Compared to the two-leg Hall ladder, the full Creutz ladder enables cross coupling between the two lattices, introduces an extra tunable magnetic flux $\phi_2$ and $\phi_3$, and provides feasibility to combine band flatness and topology. In a similar manner, the cross coupling strength is tuned by the sinusoidal RF signal amplitude and the magnetic flux is modified by the relative phases between three driving signals. 

We again set equal coupling strength within each lattice and $\phi_2=-\phi_3=\phi^{C}-\phi_A^{H}$.
In Fig.\,\ref{fig:Creutz}, b and c, We activate all three coupling types and vary the magnetic flux to experimentally reconstruct the band structures and show the related numerical calculations. Besides, when parameters are tuned to values at which the topological phase (characterized by the Zak phase) changes, the bandgap undergoes closure. Details regarding the Zak phase of the Creutz ladder and the experimental observation of this gap closing are provided in the Supplementary Information. To experimentally obtain band flatness, we adjust the DC signal to neutralize the second type of coupling $J^{V}$ and then set $J^{C}=J^{H}$, $-\phi_A^{H}=\phi_B^{H}=\pi/2$ and $\phi^{C}=0$. The obtained flat band is displayed in Fig.\,\ref{fig:Creutz}d. The Aharonov-Bohm cage effect is one of the direct topological features of the Creutz ladder under the flatness condition, where the wave function is caged within five sites although no boundary on lattices is introduced (Fig.\,\ref{fig:Creutz}e). 
We inject laser light near the center frequency (zeroth mode) into resonator A and detect the mode distribution from the drop ports of resonators A and B by scanning the length of the Fabry-Perot cavity. We observe the Aharonov-Bohm cage effect in the real (frequency) space with experimental results shown in Fig.\,\ref{fig:Creutz}, f and g. The distribution exhibits clear suppression at the zeroth site of lattice B and at higher-order sites (absolute mode index larger than 1) on both lattices, characterizing the Aharonov-Bohm cage effect. This effect is insensitive to the input laser frequency and further details are provided in the Supplementary Information. As comparison, the experimental results of the not-caged situation where $\phi_A^{H}=\phi_B^{H}=0$ (Fig.\,\ref{fig:Creutz}h) is displayed in Fig.\,\ref{fig:Creutz}, i and j. In the numerical calculations, the intensities are normalized according to the experiment.


\noindent\large\textbf{Discussion}

\normalsize
In conclusion, we design an MZI-assisted double-resonator device and experimentally demonstrate its versatility and flexibility through arbitrary tuning of the MZI using both DC and RF signals. The device effectively leverages the interplay between the frequency and spatial domains, enabling access to a wide range of coupling configurations and parameters that can be finely adjusted in principle. We observe important phenomena in the Hall ladder and Creutz ladder models, including band structures modified by synthetic magnetic fields, spin-momentum locking, and the Aharonov-Bohm cage effect. These demonstrations pave the way for simulating more complex physical models and potentially predicting novel phenomena. Furthermore, the Aharonov-Bohm cage effect may be utilized for engineering sideband multiplicity or frequency shifts.

Therefore, our design holds significant potential in building large and high-dimensional networks as both the spatial and frequency domain can be expanded to a larger scale and a higher dimension. This makes it particularly suitable for realization on integrated optics for its striking merits such as high scalability and stability \cite{dai2024non-hermit,Yun2023,bao2023}. Aside from the aforementioned models, our configuration is potential for simulating more intriguing topological models and phenomena. The modulation on the MZI leads to symmetric cross couplings in both direction along the frequency dimension and thus the off-diagonal elements in the quasimomentum-space Hamiltonian are real. It is worth noting that, we can further break this symmetry by intentionally misaligning the frequency modes of the resonators and adjust the modulation frequencies accordingly, as proposed in the Floquet topological insulator \cite{Fang2012}. Consequently, imaginary parts in the off-diagonal elements are introduced and models such as the SSH model can be realized. Generally speaking, our frequency-spatial hybrid configuration can simulate non-interacting (single-particle) effective Hamiltonians with some exceptions---since nonadjacent resonators may not be easily coupled, long-range coupling in the spatial direction may face challenges. Models include two particle interactions (such as in Hubbard models) are out of our present scope. 

Moreover, TFLN is an excellent platform that not only offers these advantages but also features high non-linearity and electro-optic coefficients, making it well-suited for our approach. However, realization may face challenges, for instance, implementing multiple modulations on a single resonator requires a long modulation region, potentially conflicting with low-loss requirements and affecting the results. Despite this challenge, TFLN processing technology is advancing rapidly, with continuous improvements in component performance, such as reduced propagation and insertion losses, and increased beam splitter precision. Consequently, we believe these challenges can be addressed in the near future, and our work may pave the way for further exploration of synthetic dimensions.

\noindent\large\textbf{Methods}

\normalsize
Our device was fabricated on an x-cut lithium niobate on insulator wafer from NANOLN, where the 400-nm lithium niobate film and a 4.7-$\mu$m $\text{SiO}_2$ layer are on a 525-$\mu$m Si substrate in order. The resonators were patterned on a Hydrogen silsesquioxane (HSQ) film with electron-beam lithography (EBL). Subsequently, 200-nm lithium niobate was etched out via an $\text{Ar}^+$-based inductively coupled plasma etching (ICP-RIE) process. The residual HSQ and the redeposition were removed by a wet etching process. 
Two layers of ultraviolet photoresist LOR5A and S1813 were spin-coated on the device and the ``bridges'' for electrodes to cross the waveguides were defined by markless lithography. Then 750-nm $\text{SiO}_2$ was deposited and followed by a lift-off process. Subsequently, 800-nm copper electrodes were plated with a similar process. Finally, the electrodes were connected to a high-frequency printed circuit board (PCB) with bonding wires. The FSR of the resonators are approximate $\Omega=2\pi \times 8.9$ GHz. The length of the modulated region is approximate 5.82 $\mu$m and the length of the entire ring is approximate 14.9 mm. The loaded Q-factor before plating electrodes was $3.7 \times 10^5$, while after plating the load Q-factor decreased to $1.7 \times 10^5$ due to metal absorption. More details can be found in Supplementary Information.

The RF signals were produced by three tunable microwave sources (AnaPico APUASYN20) synchronized by the built-in reference signals and amplified by three amplifiers (Minicircuit ZVE-3W-183+). A distributed feedback laser is injected to the device after a polarization controller through grating couplers. The output signals are collected and analyzed by either a fast oscilloscope (Keysight DCA-X 86100D) or a Fabry-Perot cavity of which the mirror reflectivity is $98\%$ and the cavity length is approximate 1.5 millimeter.

\noindent\large\textbf{Acknowledgements} 
\normalsize

This work is supported by the Innovation Program for Quantum Science and Technology (No. 2021ZD0301200), the National Natural Science Foundation of China (No. 12174370, 12174376, and 11821404, 12304546), the Youth Innovation Promotion Association of Chinese Academy of Sciences (No. 2017492), Anhui Provincial Natural Science Foundation (No. 2308085QA28), China Postdoctoral Science Foundation (No. 2023M733412). This work was partially carried out at the USTC Center for Micro and Nanoscale Research and Fabrication.

\end{document}